\documentclass[aps,prb,twocolumn,floatfix,showpacs]{revtex4}
\usepackage{graphicx}
\usepackage{amsmath,bm,amssymb,amsfonts}

\begin{document}

\title{Spin polarization in the Hubbard model with Rashba spin-orbit coupling
       on a ladder}


\author{Jos\'e A. Riera}
\affiliation{Instituto de F\'{\i}sica Rosario y Departamento de
F\'{\i}sica,\\
Universidad Nacional de Rosario, Rosario, Argentina
}

\date{\today}

\begin{abstract}
The competition between on-site Coulomb repulsion and Rashba spin-orbit
coupling (RSOC) is studied on two-leg ladders by numerical techniques. By
studying persistent currents in closed rings by exact diagonalization,
it is found that the contribution to the current due to the RSOC
$V_{SO}$,
for a fixed value of the Hubbard repulsion $U$ reaches a maximum at
intermediate values of $V_{SO}$. By increasing the repulsive
Hubbard coupling $U$, this spin-flipping current is suppressed and 
eventually it becomes opposite to the spin-conserving current. The main 
result is that the spin
accumulation defined as the relative spin polarization between the two
legs of the ladder is enhanced by $U$. Similar results for this
Hubbard-Rashba model are observed for a completely different setup in which
two halves of the ladders are connected to a voltage bias and the ensuing
time-dependent regime is studied by the density matrix-renormalization
group technique. It is also interesting a combined effect between $V_{SO}$
and $U$ leading to a strong enhancement of antiferromagnetic order which
in turn may explain the observed behavior of the spin-flipping current.
The implications of this enhancement of the spin-Hall effect with electron
correlations for spintronic devices is discussed.
\end{abstract}


\maketitle

\section{Introduction}

Recent trends in the field of spintronics\cite{wolf,prinz,zutic} exploit
the possibility of controlling electron spins by purely electrical
means without using magnetic materials,\cite{awschalom} which is at
the heart of the conceptual proposal of the Datta-Das spin field-effect
transistor.\cite{dattadas} Key to this possibility is the Rashba
spin-orbit (SO) interaction\cite{rashba,winkler}
appearing due to structure inversion asymmetry of materials and which has 
been shown to be tuned by gate voltages in semiconductor 
heterostructures.\cite{miller}

Although the main interest in the study of RSOC has been on its
ability of generating and controlling spin polarized currents,
another consequence of the interplay between
charge and spin degrees of freedom induced by the RSOC is a flow of
spins transversal to the flow of charge which is known as the
spin-Hall effect.\cite{dyakonovperel,Hirsch99,murakami,sinova04} If
transport takes place along a planar strip, the spin-Hall effect leads
to a spin accumulation with opposite projection on both edges of the
strip.\cite{nikolic,malshukov} This has been experimentally
observed.\cite{kato_spin_acc,wunderlich}
Hence, the spin-Hall effect provides a real-space separation between
up and down spin electrons, like two ferromagnets with
opposite polarizations, thus enhancing the metallic state and its
conductance, and these features could also be employed in spintronics
devices.

Since most devices and experiments about RSOC have involved
semiconductors, electron correlations have mostly not been included so
far in theoretical models.
However, there are some systems which have been recently considered
in this field where such correlations may play an important role.
In the first place, RSOC has been observed in oxide
heterostructures.\cite{caviglia,nakamura,pdcking} Important
effects of a RSOC have been experimentally observed in other
transition metal oxides, particularly the iridates where the relevant
degrees of freedom are the 5d electrons on the Ir$^{4+}$ ions, and
where strong correlation effects are expected.\cite{bjkim,trescher}
It should be also taken into account that correlation
effects become increasingly important in systems with low
spatial dimensions. Interesting crossovers have been reported when
moving from two-dimensional (2D) systems to quantum wires.\cite{wenk}
In this sense, spin-orbit effects, predominantly of the Rashba type
have been studied in quantum wires.\cite{quay,hansen}
The presence of RSOC in graphene and carbon nanotubes has been 
pointed out\cite{huertas} and in fact a theoretical study with up
to two interacting electrons has been done on a nanotube quantum
dot.\cite{wunsch}

In this work, an attempt of studying the interplay between electron
correlations and RSOC will be performed by considering a Hubbard model
with a Rashba SOC. During the last two decades it has been realized the
difficulties of tackling electron correlations theoretically. In studies
of RSOC, electron correlations have been included only locally regarding
quantum dots\cite{qfsun0,mross} or on effective onedimensional (1D)
systems using Tomonaga-Luttinger liquid approach\cite{pletyukhov}
and conformal field theory.\cite{zvyagin} For
this reason, in the present work a two-leg ladder, which may be the
narrowest quasi-1D geometry on which a spin-Hall effect
could take place will be considered. The Hubbard-Rashba model on a 
two-leg ladder is suitable for treatment using essentially exact,
unbiased, computational techniques, which have been enormously 
valuable in the field of strongly correlated electron systems.
Particularly for mesoscopic systems, the full quantum many-body
interplay between different terms of the Hamiltonian can be
reliably captured. There are many compounds that contain two-leg
ladder structures such as cuprates\cite{azuma,uehara}, 
vanadates\cite{smirnov}, iron chalcogenides\cite{caron}, and other
organic and inorganic compounds, although so far the presence of a 
SOC has not been reported on such materials. It should be also
stressed that a Hubbard model in a two-leg ladder has been proposed
for carbon nanotubes.\cite{balentsfisher}
Then, the main goal is to determine and understand
the behavior of the spin accumulation as a function of the RSOC and
the Hubbard repulsion $U$. The interplay between RSOC and $U$ with
respect to a possible metal-insulator transition\cite{pesin} will not
be discussed in the present work.

\begin{figure}
\includegraphics[width=\columnwidth,angle=0]{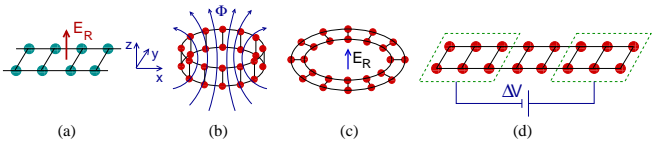}
\caption{(Color online) (a) The two-leg ladder system here considered;
(b) a two-leg ring pierced by a magnetic flux; (c) previously study 2D
rings; (d) setup used in DMRG calculations.}
\label{fig1}
\end{figure}

\section{Model and methods}

The Hamiltonian of the Hubbard-Rashba model in a
square lattice is defined as:
\begin{eqnarray}
H &=& - t_0 \sum_{<l,m>,\sigma} (c_{l,\sigma}^\dagger c_{m,\sigma} +
H. c.) + U \sum_{l} n_{l,\uparrow} n_{l,\downarrow} \nonumber  \\
&+& V_{SO} \sum_{l} [c_{l+x,\downarrow}^\dagger c_{l,\uparrow} -
c_{l+x,\uparrow}^\dagger c_{l,\downarrow} + i (
c_{l+y,\downarrow}^\dagger c_{l,\uparrow} \nonumber  \\
&+& c_{l+y,\uparrow}^\dagger c_{l,\downarrow}) + H. c.]
\label{hamtot}
\end{eqnarray}
where the notation is standard. The first two terms corresponds to the 
usual Hubbard
model and the last term contains the RSOC.\cite{pareek}
The couplings $t_0$ and $V_{SO}$ are normalized in such a way
that $t_0^2+V_{SO}^2=1$ (which is henceforth taken as the unit of energy). 
The longitudinal (transversal) direction of the ladder corresponds to the
$x$-axis ($y$-axis), as shown in Fig.~\ref{fig1}(a). The RSOC in
(\ref{hamtot}) corresponds to an effective Rashba electric field $E_R$ 
along the $z$-axis, i.e. perpendicular to the plane of the ladder. Open
boundary conditions (OBC) are applied in the transversal direction. The
total number of lattice sites is $N=2 \times L$. Hamiltonian 
(\ref{hamtot}) corresponds to the isotropic ladder, in which couplings
are the same along both directions.
The present study is limited to the quarter filled system ($n=0.5$).
Weak-coupling analysis\cite{balents} has shown that the Hubbard
model on the isotropic ladder at $n=0.5$ is at the boundary between
a metal (C1S0) and an insulating (C0S1) phase.\cite{rpd} Exactly at 
the boundary,
where additional metallic phases are predicted to be stable, both this 
study as well as numerical studies\cite{noack,vojta} and the present
results point to a metallic phase, even for the values of $U>0$ 
here examined. As soon as $V_{SO}$ is turned on, the occupied 
bonding band will split leading to a four-point Fermi surface
and the system will behave as a metal.
In all cases, the length $L$, equal to the total number of electrons,
was set equal to $4 m$ ($m$ integer).

In order to induce persistent currents, a magnetic flux piercing the
ring formed by a closed ladder has to be applied (Fig.~\ref{fig1}(b)). 
This magnetic flux $\Phi$ is included in the Hamiltonian by the usual
Peierls factors, which in momentum space is equivalent to replacing
$k_x$ by $k_x+\Phi/L$ (units
in which $e=c=\hbar=1$ have been adopted). These Peierls factors are
included in both the hopping and SO terms of (\ref{hamtot}). The 
resulting Hamiltonian is studied using exact diagonalization (ED)
for $U> 0$. For the noninteracting case, $U=0$, by working in
momentum space, ladders with $L$ up to 6400 have been
considered.
Due to the RSOC the total $S^z$ is not conserved hence all possible
values of total $S^z$ have to be included in the Hilbert
space. In order to reduce the Hilbert space dimension, translation
symmetry along the longitudinal direction has been implemented thanks
to the use of periodic boundary conditions (PBC) along the $x$-axis.
Notice that the ladder ring shown in Fig.~\ref{fig1}(b)
is different from the 2D ring (Fig.~\ref{fig1}(c)) that has been
intensively studied in the
non-interacting case.\cite{splettstoesser,XinLiu,shengchang} Studies
including Coulomb interactions for these rings have been limited to few
electrons.\cite{daday}

The total current is defined as
$J(\Phi)=\frac{\partial E_0(\Phi)}{\partial \Phi}$
where $E_0(\Phi)$ is the ground state energy. The contribution from
the hopping term in (\ref{hamtot}), the hopping or spin-conserving
current, is then given by the usual current operator along the legs,
$\hat{j}_h = i t_0 \sum_{l,\sigma} (c_{l+\hat{x},\sigma}^\dagger
c_{l,\sigma} - H. c.)$. A similar current operator can be easily
derived for the contribution from the RSOC term in (\ref{hamtot}),
that is, the SO or spin-flipping current, $\hat{j}_{SO}$. Notice
that $j_{SO}$ turns out to be proportional to $V_{SO}$. Using the
Feynman-Hellmann theorem, the total current is given by
$L J=j_h+j_{SO}$. In ED calculations, $J$, $j_h$, and $j_{SO}$ have
been independently computed.
The spin polarization due to the spin-Hall effect
is defined as $\Delta S^z = <S^z_1 - S^z_2>/2$, where $S^z_j$ is the
total z-component of the spin on leg $j$. All physical quantities are
ground state averages, and hence, they are functions of the magnetic
flux $\Phi$. In order to determine the effects of the Rashba and 
Hubbard couplings on these quantities, the reasonable choice of looking
at their maximum values as a function of the flux $\Phi$ was adopted.

The second setup considered in this work is shown in Fig.~\ref{fig1}(d),
which corresponds to OBC along the $x$-axis. Here, at a given time the
two ends of a ladder are
connected to a voltage bias $\Delta V$ generating a time-dependent
regime which is studied by DMRG. Although various sophisticated
approaches have been proposed within DMRG to deal with this time 
evolution,\cite{feiguinwhite,schollwock,schmitteckert} the simple
``static" approach is implemented in the present work which is
enough to reliably capture the main physical features in the
small lattices studied. This is a non-equilibrium process, but a
small value $\Delta V=0.01$ was adopted. In this case, the total
current is computed as $J(t)=\frac{\partial N_l}{\partial t}$, where
$N_l$ is the total charge in the left half of the ladder, and $t$ is
the time measured in units of the time increment $\Delta \tau=0.1$.
$J_{max}$ is computed as the average of $J(t)$ between the first two
peaks of its time evolution, which can be precisely determined with a 
relatively small number of retained states $M$ within the static 
approach. The spin-conserving current
$j_h(t)$ was computed at the central links of the ladder. $j_{h,max}$
was computed using the same criterion as for $J_{max}$.
In DMRG calculations, $J$ and $j_h$ have been independently computed.

\section{Results}

\subsection{Noninteracting case}

The tight-binding part of the Hamiltonian is,
in momentum space:
\begin{eqnarray}
H_0=\sum_{\mathbf{k}}
\begin{pmatrix}
A & B & -t_\perp & -i V_{SO,\perp} \\
B^* & A & -i V_{SO,\perp} & -t_\perp \\
-t_\perp & i V_{SO,\perp} & A & B \\
i V_{SO,\perp} & -t_\perp & B^* & A
\end{pmatrix} \qquad
\label{tigh_binding}
\end{eqnarray}
in the basis: $(c_{1,\mathbf{k},\uparrow}, c_{1,\mathbf{k},\downarrow,} 
c_{2,\mathbf{k},\uparrow}, c_{2,\mathbf{k},\downarrow,})$, where ``1",
and ``2" are the two legs of the ladder.

\begin{figure}
\includegraphics[width=0.43\textwidth]{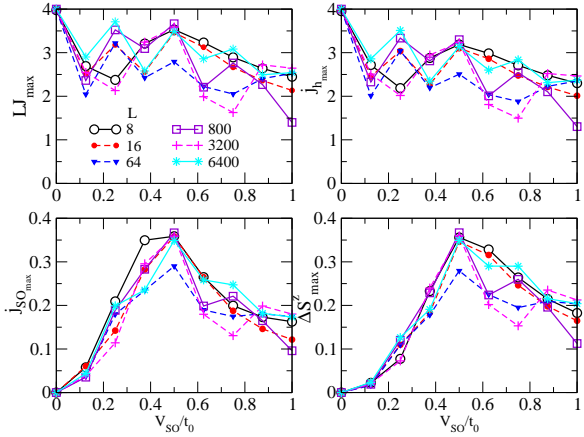}
\caption{(Color online) Maximum value of (a) total current, (b)
hopping current, (c) spin-orbit current, and (d) spin accumulation
$\Delta S^z$ as a function of $V_{SO}/t_0$ for the noninteracting system
on various two-leg ladders,
$n=0.5$. Tight-binding calculations for the persistent currents setup.}
\label{fig2}
\end{figure}

For the general spatially anisotropic ladder:
\begin{eqnarray}
A &\equiv& -2 t_\parallel \cos{k_x}  \nonumber  \\
B &\equiv& -2 i V_{SO,\parallel} \sin{k_x}
\end{eqnarray}
In the isotropic case here considered,
$t_\parallel = t_\perp =t_0$, $V_{SO,\parallel} = V_{SO,\perp} =V_{SO}$.

In the tight-binding formalism, the currents $J$, $j_h$, and $j_{SO}$
are independently computed for each value of the magnetic flux $\Phi$,
and hence the relation $LJ=j_h+j_{SO}$ provides an internal check of
the calculations.

\begin{figure}
\includegraphics[width=0.83\columnwidth,angle=0]{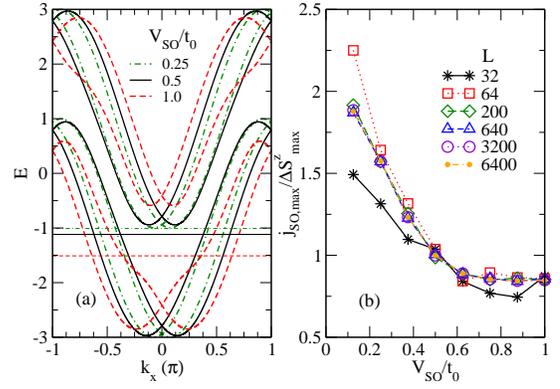}
\caption{(Color online) (a) Band structure for various values of
$V_{SO}/t_0$.
The chemical potentials are indicated with the same convention.
(b) Ratio of the maximum values of spin flipping current to the 
spin accumulation as a function of $V_{SO}/t_0$. Length of the
ladder rings are indicated in the plot.}
\label{fig1_sm}
\end{figure}

For the case of persistent currents, Fig.~\ref{fig1}(b), results for
the noninteracting $U=0$ case are shown in Fig.~\ref{fig2} for
various lattice sizes ranging from $L=8$ to 6400. For the total and
the hopping currents ((Fig.~\ref{fig2}(a),(b)), a general trend to 
decrease with increasing $V_{SO}/t_0$ can be observed, although there
is a considerable noise due to various level crossings as a function
of the flux $\Phi$. In addition it can be observed that
the maximum values of the spin-flipping current (Fig.~\ref{fig2}(c))
and the spin accumulation (Fig.~\ref{fig2}(d)) occur at 
$V_{SO}/t_0\approx 0.5$. The general decrease of $J_{max}$ and
$j_{h,max}$ can be explained by a shift of the chemical potential
away from the bottom of the anti-bonding bands (Fig.~\ref{fig1_sm}(a)).
The behavior of $j_{SO,max}$ is more complex and it is determined
by the competition between the fact that $V_{SO}$ is an overall factor
in the operator $\hat{j}_{SO}$, and the increase in the shift between 
the upper and lower spin bands (Fig.~\ref{fig1_sm}(a)) with $V_{SO}$,
which works against spin-flipping processes. For $V_{SO}/t_0 < 0.5$
the band structure does not change too much and $j_{SO,max}$
increases approximately linearly with $V_{SO}/t_0$. For
$V_{SO}/t_0 > 0.5$, the band structure changes more strongly and
$j_{SO,max}$ starts to be suppressed. The behavior of 
$\Delta S^z_{max}$ follows that of $j_{SO,max}$, and it is remarkable
that in spite of the noise introduced by level crossings, the 
ratio $j_{SO,max}/\Delta S^z_{max}$ follows a smooth monotonous
behavior as a function of $V_{SO}/t_0$ for large $L$, as shown in
Fig.~\ref{fig1_sm}(b).

\begin{figure}
\includegraphics[width=0.43\textwidth]{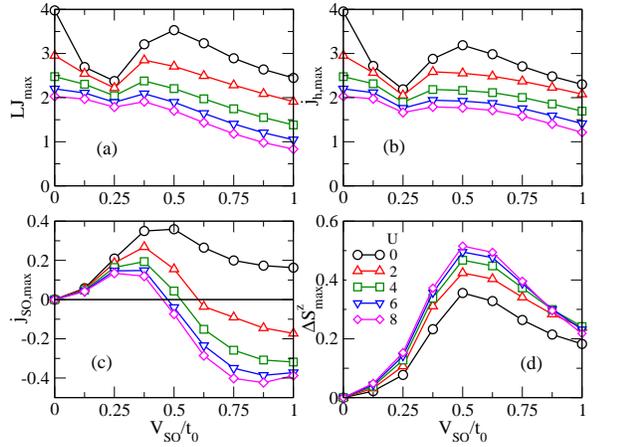}
\caption{(Color online) Maximum value of (a) total current, (b) hopping
current, (c) spin-orbit current, and (d) spin accumulation $\Delta S^z$ 
as a function of $V_{SO}/t_0$ for various values of $U$. Results for 
the $2 \times 8$ ladder, $n=0.5$, obtained with ED.
}
\label{fig3}
\vspace{-0.3cm}
\end{figure}

\subsection{Interacting case}

Let us discuss the correlation effects for the case of persistent 
currents, for $0 \leq U \leq 8$. Fig.~\ref{fig3} shows the results 
for the maximum values of the total
current, $J$, the hopping and spin-orbit contributions to the current,
$j_h$ and $j_{SO}$, and the spin-leg polarization $\Delta S^z$ 
on the $2 \times 8$ ladder at $n=0.5$. As it can be seen in
Fig.~\ref{fig3}(a), the total current for a fixed value of 
$V_{SO}/t_0$ decreases with increasing $U$. This result is expected 
for $V_{SO}=0$, i.e. for the pure Hubbard model, where it is well known
that the conductance  $G$ decreases with increasing $U$, as in the 
one-dimensional Hubbard model presenting a Luttinger liquid behavior
where $G\sim K_\rho$,\cite{kanefisher} where $K_\rho$ is the 
Luttinger exponent that determines the long-range distance 
behavior of correlations. This behavior was examined for ladders in
Ref.~\onlinecite{orignac}. The maximum value of the hopping current,
shown in 
Fig.~\ref{fig3}(b), follows roughly the same behavior with $U$ and
$V_{SO}/t_0$ as the total current since it is much larger than the 
spin-flipping contribution. The dip in $J$ and $j_h$ at 
$V_{SO}/t_0\approx 0.25$ (Fig.~\ref{fig3}(a),(b)) is just a finite
size effect as discussed above. On the
other hand, the behavior of the maximum value of the SO current is
quite interesting (Fig.~\ref{fig3}(c)). For all values of $U$, 
$j_{SO,max}$ increases with $V_{SO}/t_0$ until reaching a
maximum which depends on $U$, and then starts to decrease becoming
eventually negative, indicating that its direction is {\em opposed}
to the hopping current, by convention. The most important results 
correspond to the maximum value of the spin-leg polarization shown
in Fig.~\ref{fig3}(d). One should first notice that $\Delta S^z_{max}$
increases as a function of the Coulomb repulsion $U$ for all values of
$V_{SO}/t_0$. In addition, its maximum value near $V_{SO}/t_0=0.5$ 
observed for the noninteracting case is preserved for finite $U$.

\begin{figure}
\includegraphics[width=0.85\columnwidth,angle=0]{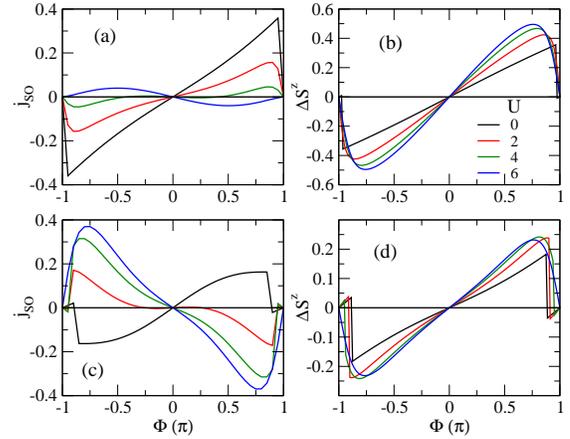}
\caption{(Color online) (a) Spin flipping current, and (b) spins
 accumulation, as a
function of the magnetic flux for various values of $U$, at 
$V_{SO}/t_0=0.5$.  (c) Spin flipping current, and (d) spin 
accumulation, as a function of the magnetic flux for various values 
of $U$, at $V_{SO}/t_0=1$. Results for the $2 \times 8$ ladder.}
\label{fig2_sm}
\end{figure}

The dependence of $j_{SO}$ and $\Delta S^z$ as a function of the
magnetic flux, for $V_{SO}/t_0=0.5$ and 1, is shown in 
Fig.~\ref{fig2_sm}. Notice that increasing $U$ produces a smooth 
change in the overall shape of the curve of the flux dependence of
the computed quantities, thus justifying the adopted criterion of
analyzing their maximum values as a function of $U$ and $V_{SO}/t_0$.
In particular, what is interesting in this plot is that the 
spin-flipping current $j_{SO}$ (Fig.~\ref{fig2_sm}(a)) is inverted
for $U>4$ for $V_{SO}/t_0=0.5$ thus becoming opposite to the 
spin-conserving current $j_h$, as discussed before. For 
$V_{SO}/t_0=1$, this feature is observed already for $U=4$
Fig.~\ref{fig2_sm}(b).

\begin{figure}
\vspace{0.6cm}
\includegraphics[width=0.43\textwidth]{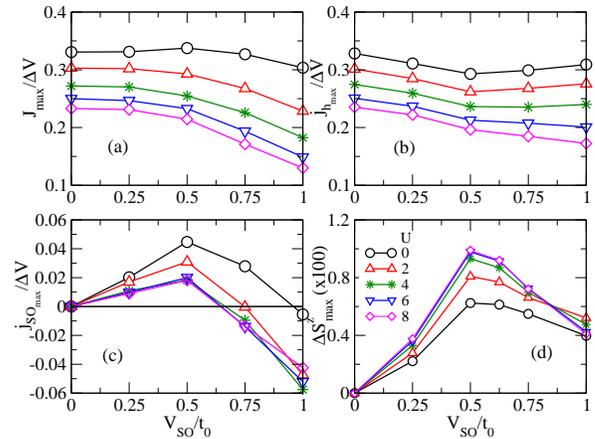}
\caption{(Color online) Maximum value of (a) total, (b) hopping, and
(c) spin-orbit currents, divided by $\Delta V$, and (d) spin accumulation
$\Delta S^z$ as a function of $V_{SO}/t_0$ for various
values of $U$.  Results for the $2 \times 20$ ladder, $n=0.5$,
obtained with DMRG.
}
\label{fig4}
\end{figure}
Results for the voltage bias setup (Fig.~\ref{fig1}(d)), for the 
$2 \times 20$ ladder, are shown in
Fig.~\ref{fig4}. The behavior of the maximum values of the total, 
spin-conserving and spin-flipping currents, and of the spin accumulation
as a function of $V_{SO}/t_0$ for various values of $U$, are qualitatively
strikingly similar to those shown for the setup involving persistent 
currents (Fig.~\ref{fig3}) in spite of the quite different background
physics involved. Notice that actually the maximum values of the currents
divided by the voltage bias $\Delta V$ are shown, which correspond to the
conductance $G$ up to a factor of $2 \pi$. Comparison with results for
the $2 \times 16$ do not show appreciable finite-size effects.

\begin{figure}
\vspace{0.6cm}
\includegraphics[width=0.43\textwidth,angle=0]{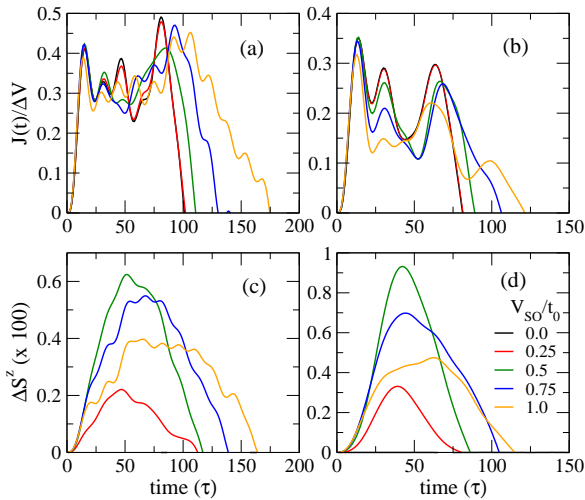}
\caption{(Color online) Time evolution of the total current $J$ divided by 
$\Delta V$ for (a) $U=0$ and (b) $U=4$, and time evolution of the spin 
accumulation for (c) $U=0$ and (d) $U=4$, for various values of
$V_{SO}/t_0$ indicated on the plot. Results for the $2 \times 20$
ladder, obtained by DMRG.
Only the first half-period of the time evolution is shown, 
for the next half-period both $J$ and $\Delta S^s$
reverse their sign.}
\label{fig4_sm}
\end{figure}

In Fig.~\ref{fig4_sm} some typical time evolution of the total
current and of the spin accumulation are shown for various values of
$U$ and $V_{SO}/t_0$. It is well-known that the time evolution of
the current follows an oscillatory behavior due to the interplay
between kinetic and potential energies. As the number of retained
states increases, or by adopting an adaptive scheme\cite{schollwock}
in time-dependent DMRG, the current reaches a smooth plateau during
each half-period of the time evolution.  The value of the current at
the plateau divided by the voltage bias $\Delta V$ is equal $G/2\pi$,
where $G$ is the conductance (remember that units where $e=\hbar=1$
were adopted). In the present case, the adopted criteria of
measuring $J_{max}$ as the average of $J(t)$ between the first
two peaks leads to a slight overestimation of $G$ with respect to its
exact value at $U=V_{SO}=0$.

\begin{figure}
\vspace{0.4cm}
\includegraphics[width=0.43\textwidth]{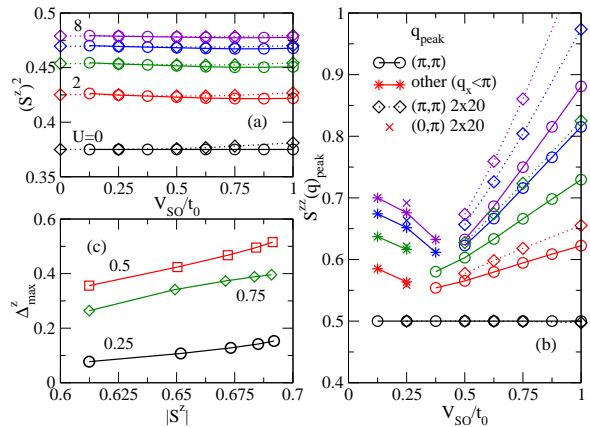}
\caption{(Color online) (a) $(S^z)^2$ as a function of $V_{SO}/t_0$
for $U=0, 2, 4, 6, 8$ from bottom to top for the $2 \times 8$ ladder
PBC at zero flux (circles) and for the $2 \times 20$ ladder OBC
(diamonds). (b) Peak value of the magnetic structure factor
$S^{zz}({\bf q})$ as a function of $V_{SO}/t_0$ for $U=0, 2, 4, 6, 8$
from bottom to top for the $2 \times 8$ ladder PBC at zero flux
(circles, stars) and for the $2 \times 20$ ladder OBC (diamonds,
crosses) if ${\bf q}=(\pi,\pi)$ (other) 
(c) Maximum value of the spin accumulation $\Delta S^z$ as a function
of $|S^z|$ at zero flux for various values of $V_{SO}/t_0$ indicated
in the plot, for the $2 \times 8$ ladder PBC.
}
\label{fig5}
\end{figure}

\subsection{Magnetic properties}

Some possible explanations of the behaviors shown in Figs.~\ref{fig3}
and \ref{fig4} can be obtained by examining the dependence
of various magnetic properties with the parameters $U$ and $V_{SO}/t_0$.
$(S^z)^2$ and the static magnetic structure factor $S({\bf q})$ are 
defined as usual from the $z-z$ spin correlations:
\begin{eqnarray}
(S^z)^2=\frac{1}{N}\sum_{j,l} <(S^z_{j,l})^2> \nonumber \\
S({\bf q})=\frac{1}{N}\sum_{\bf r_1,r_2}<S^z_{\bf r_1}
S^z_{\bf r_2}>e^{i\bf q \cdot (r_1-r_2)}
\label{spineqs}
\end{eqnarray}
where the normalization $S^z_{j,l}$ equal $+1$ ($-1$) for up (down)
spin projections was used. With this definition, it is easy to find 
that for the non-interacting case 
$(S^z)^2=n_\uparrow (1-n_\downarrow)+ (\uparrow \leftrightarrow 
\downarrow)$, where $n_\uparrow$ ($n_\downarrow$) is the density
of electrons with $\uparrow$ ($\downarrow$) spin in a state with
fixed total $S_{total}^z$. For $S_{total}^z=0$, $(S^z)^2=0.375$,
and when $U \rightarrow \infty$, $(S^z)^2 \rightarrow n=0.5$,
since doubly-occupied sites become forbidden.

In Fig.~\ref{fig5}(a) the average value of $(S^z)^2$, a measure of the
magnetic moment per site as a function of 
$V_{SO}/t_0$ and for various values of $U$ is shown. As expected,
$(S^z)^2$ increases with $U$, as it is well-known for Hubbard-like 
models, and it is almost independent of $V_{SO}/t_0$. Results for
the $2 \times 8$ ladder PBC at zero flux are virtually indistinguishable
with the ones obtained for the $2 \times 20$ ladder OBC using DMRG.
Note that for $U=0$, one gets the same value as for $S_{total}^z=0$,
in spite of the fact that since $S_{total}^z$ is not a quantum number
for the model here considered, the ground state is a combination of
subspaces with all possible $S_{total}^z$.
More interesting are the results for the peak value of the static
magnetic structure factor  $S^{zz}({\bf q})$, which are shown in 
Fig.~\ref{fig5}(b) as a function of $V_{SO}/t_0$ and for various values 
of $U$. For large values of $V_{SO}/t_0$, $S^{zz}({\bf q})$ has a peak
at ${\bf q}=(\pi,\pi)$, while for small values of $V_{SO}/t_0$,
$S^{zz}({\bf q})$ is maximal at other values of ${\bf q}$, all of
them with $q_x \neq \pi$, and in most cases at ${\bf q}=(0,\pi)$. In
the first place, as in Fig.~\ref{fig5}(a), it is expected that
$S^{zz}({\bf q})$ is enhanced by $U$. What is unexpected is the
strong dependence of $S^{zz}({\bf q})$ with $V_{SO}/t_0$:
$S^{zz}(\pi,\pi)$ increases with increasing $V_{SO}/t_0$, and
$S^{zz}({\bf q}\neq (\pi,\pi))$ increases with decreasing $V_{SO}/t_0$.
This is a combined effect of $V_{SO}/t_0$ and $U$ since for $U=0$,
$S^{zz}({\bf q})$ is constant and clearly the slope of
$S^{zz}({\bf q})$ with $V_{SO}/t_0$ increases with $U$. Results for
the $2 \times 20$ ladder show a stronger dependence than the ones
corresponding to the $2 \times 8$ ladder and this could be explained
by the fact that in DMRG $S^{zz}({\bf q})$ calculations, the only
included spin-spin correlations are the ones
measured from one of the central sites of the ladder.

The increase of the z-component of the magnetic moment can at least 
partially explain the larger values of the spin accumulation observed
as a function of the Hubbard repulsion for a fixed value of $V_{SO}/t_0$.
This is shown in Fig.~\ref{fig5}(c) where it can be observed an almost 
linear dependence of $\Delta S^z_{max}$ with $|S^z|$ which is expected
for an density operator. Of course, it is not trivial the dependence
of the slope and $U=0$ values with $V_{SO}/t_0$.
The peak of the magnetic structure factor at
${\bf q}=(\pi,\pi)$, indicating the presence of an antiferromagnetic
(AFM) order could in turn
provide some insight on the suppression of the spin-flipping current
$j_{SO}$ for large $V_{SO}/t_0$ and $U$, shown in Figs.~\ref{fig3}(c)
and \ref{fig4}(c). For an AFM order, two electrons occupying nearest
neighbors sites along one leg of the ladder would likely have spins
with opposite projections on the $z$-axis. For this configuration,
a spin-flipping hopping would be impossible while there would be no 
problem for a spin-conserving hopping (except for a cost $~U$). Now,
once a $t_0$-hopping has occurred, creating a doubly-occupied site
and leaving behind an empty site, a $V_{SO}$-hopping is now possible
from the doubly-occupied 
site to the empty one. Hence the spin-flipping hopping takes place in
a direction {\em opposite} to the spin-conserving one thus possibly
explaining the observed change of sign of the spin-flipping current.

\section{Conclusions}

To summarize, we have considered perhaps the
simplest system where some features associated to Rashba spin-orbit
coupling can be observed.\cite{malshukov} This two-leg ladder
system is suitable to be studied by unbiased, essentially
exact, computational techniques, which allow to consider electron 
correlations on an equal foot with spin-conserving and spin-flipping
hoppings. At quarter-filling it was observed that spin accumulation
is maximal at an intermediate value of $V_{SO}/t_0$, independently of
$U$. By increasing a repulsive $U$, the spin-flipping current is
suppressed and it becomes negative for large $V_{SO}$ and $U$, while 
spin accumulation is enhanced. These two last features can be 
explained by the increase of $(S^z)^2$ and AFM correlations,
the latter being a combined effect of $V_{SO}$ and $U$. The present
approach could be extended to wider strips which could in principle
be studied by DMRG, and also by variational and diffusion Monte Carlo
(plus a fixed-node approximation), and to 
different electron fillings.
In addition, some subtle and so far controversial properties, such
as spin currents, could be studied numerically on this class of strips
including electron correlations. Although the present study does not 
include any dissipative mechanism, an enhancement of spin accumulation
by electron interactions together with a reduction of the overall
current are positive features that could be taken advantage in 
spintronic devices.

The author wishes to acknowledge Sadamichi Maekawa for calling his
attention to related problems and to George Martins for a careful
reading of the manuscript. The author is supported in part by the
Consejo Nacional de Investigaciones Cient\'{\i}ficas y T\'ecnicas,
Argentina.

\end{document}